\documentclass[fleqn,10pt,amsmath,amssymb]{wlscirep}
\usepackage[utf8]{inputenc}
\usepackage[T1]{fontenc}

\usepackage{epsfig,graphicx,color}
\usepackage{booktabs,siunitx}
\usepackage{float}   
\usepackage{subcaption}
\usepackage{float}

\usepackage{tabularx}
\newcolumntype{Y}{>{\raggedleft\arraybackslash}X}

\newcommand{\bp}{{\bf p}}
\newcommand{\bpp}{\bp'}
\newcommand{\bpz}{\bp''}

\newcommand{\bphat}{\hat{\bf p}}

\newcommand{\lam}{\lambda}
\newcommand{\Vr}{{\cal V}}

\newcommand{\brap}{\ ^{\pi a}\langle \bp; \bphat s \lam ; t |}

\newcommand{\brappleftprime}{\ ^{\pi' a}\langle \bp';\bphat' s' \lam' ; t' }

\newcommand{\ketp}{| \bp; \bphat s \lam ; t \rangle^{\pi a}}
\newcommand{\ketpp}{| \bp' ; \bphat' s \lam' ; t \rangle^{\pi a}}

\newcommand{\pist}{{\pi s t}}
\newcommand{\pizt}{{\pi 0 t}}
\newcommand{\piot}{{\pi 1 t}}

\newcommand{\pistlp}{{\pi s t, \lambda'}}
\newcommand{\piotlp}{{\pi 1 t, \lambda'}}

\newcommand{\piztz}{{\pi 0 t, 0}}

\newcommand{\omdp}{\omega(p)  +   \omega(p')}

\title{Relativistic nucleon-nucleon potentials in a spin-dependent three-dimensional approach}

\author[1,2,$\dagger$]{M. R. Hadizadeh}
\author[3,$\ddagger$]{M. Radin}
\author[3,$\ast$]{F. Nazari}
\affil[1]{College of Engineering, Science, Technology and Agriculture, Central State University, Wilberforce, OH
45384, USA,}
\affil[2]{Department of Physics and Astronomy, Ohio University, Athens, OH 45701, USA,}
\affil[3]{Department of Physics, K. N. Toosi University of Technology, Tehran, Iran.}

\affil[$\dagger$]{mhadizadeh@centralstate.edu}
\affil[$\ddagger$]{radin@kntu.ac.ir}
\affil[$\ast$]{f.nazari@email.kntu.ac.ir}

\keywords{Relativistic nucleon-nucleon potentials, Three-dimensional scheme, Lippmann-Schwinger equation}

\begin{abstract}
The matrix elements of relativistic nucleon-nucleon $(NN)$ potentials are calculated directly from the nonrelativistic potentials as a function of relative $NN$ momentum vectors, without using a partial wave decomposition. To this aim, the quadratic operator relation between the relativistic and nonrelativistic $NN$ potentials is formulated in momentum-helicity basis states. It leads to a single integral equation for the two-nucleon $(2N)$ spin-singlet state and four coupled integral equations for two-nucleon spin-triplet states, which are solved by an iterative method. Our numerical analysis indicates that the relativistic $NN$ potential obtained using CD-Bonn potential reproduces the deuteron binding energy and neutron-proton elastic scattering differential and total cross-sections with high accuracy.
\end{abstract}

\begin{document}

\flushbottom
\maketitle
\thispagestyle{empty}

\section{Introduction}
\label{Introduction}
Since Einstein's theory of special relativity in the early 20th century, there are still issues when considering systems that contain more than two nucleons, as in such systems, pair nucleons are influenced by the presence and motion of the other nucleons. As the three- and four-nucleon bound and scattering problems can be numerically solved with controlled errors, they provide an ideal theoretical laboratory for investigating the relativistic effects in the few-nucleon systems. Several techniques are developed to study the relativistic effects in few-body systems, and among them, the Faddeev-Yakubovsky method provides an exact numerical treatment of three and four-nucleon systems.

The inputs for the relativistic Faddeev-Yakubovsky equations are the fully-off-shell (FOS) relativistic $2N$ $t-$matrices, which can be obtained with two different approaches. In the first approach, the FOS relativistic $2N$ $t-$matrices are obtained directly from the nonrelativistic $2N$ $t-$matrices applying a two-step process. In the first step, using an analytical relation proposed by Coester {\it et al.}, the relativistic right-half-shell (RHS) $t-$matrices are obtained from the nonrelativistic RHS $t-$matrices \cite{Coester1975}. In the second step, the FOS relativistic $t-$matrices are obtained from the RHS $t-$matrices by solving a first resolvent equation proposed by Keister {\it et al.} \cite{Keister2006}.
This approach has been successfully implemented in few-body bound and scattering calculations in a three-dimensional (3D) scheme \cite{hadizadeh2016relativistic,hadizadeh2014relativistic,lin2008relativistic,elster2007relativistic,lin2007first,liu2007three}, without using a partial wave (PW) decomposition.

In a second approach, the relativistic FOS $t-$matrices can be calculated from the solution of the relativistic Lippmann-Schwinger (LS) equation using the relativistic $2N$ potentials. The input relativistic $NN$ potentials can be obtained from the nonrelativistic potentials by solving a quadratic equation, using an iterative scheme proposed by Kamada and Gl\"ockle \cite{kamada2007realistic}. We have recently implemented this iterative technique in a 3D scheme to calculate the matrix elements of relativistic two-body (2B) potentials for spin-independent Malfliet-Tjon (MT) potential as a function of the magnitude of 2B relative momenta and the angle between them. To do so, we formulated the quadratic operator relation between the nonrelativistic and relativistic $NN$ potentials in momentum space leading to a 3D integral equation \cite{Hadizadeh2017}. We successfully implemented this iterative approach to calculate the matrix elements of boosted 2B potential from the MT potential to study the relativistic effects in a 3B bound state \cite{Hadizadeh2020}. Our numerical results showed that the relativistic effects lead to a 2\% reduction in 3B binding energy using MT potential.
Our exact and detailed numerical studies of relativistic effects in 3B bound states using spin-independent MT potential demonstrates that direct integrations in the 3D scheme can be utilized to achieve the same results obtained using a PW method and paves the path for an extension to realistic interactions that have a more complicated spin-isospin dependence.
Considering modern nucleon-nucleon ($NN$) potentials by including spin and isospin degrees of freedom and calculating the relativistic $NN$ potentials from realistic $NN$ potentials is the task we address in this paper. 
%
This is the first step toward our goal for a fully relativistic treatment of triton and Helium-3 bound state properties and the long-term interest in studying the scattering problems at the few-GeV energy scale in a 3D scheme. 
We show that the representation of the quadratic equation in momentum helicity basis states leads to a single and four coupled 3D integral equations for $NN$ singlet and triplet spin state, correspondingly. The single and coupled integral equations are solved using the mentioned iterative scheme, and the matrix elements of relativistic $NN$ potentials are obtained from CD-Bonn potential \cite{machleidt2001high}.
Our numerical analysis indicates that the calculated relativistic potential reproduces the deuteron binding energy and differential and total cross-sections of neutron-proton ($np$) elastic scattering with very high accuracy.

The motivation for using the 3D scheme and implementing a direct integration method is to replace the discrete angular momentum quantum numbers of a PW representation with continuous angle variables and consider all partial wave components to infinite order, independent of the energy scale of the problem. Consequently, 3D representation avoids the very involved angular momentum algebra for permutations, transformations, and few-nucleon forces, and in contrast to the PW approach, the number of equations in the 3D representation is energy independent. 
The 3D scheme is successfully implemented in a series of few–body bound and scattering states calculations by different few-body groups, from Ohio-Bochum collaboration 
\cite{elster1998two,
elster1999three,
schadow2000three,
fachruddin2000nucleon,
fachruddin2001new,
fachruddin2003n,
liu2003model,
Fachruddin_PRC69,
liu2005three,
lin2007first,
lin2008relativistic,
lin2008poincare,
glockle2010new,
veerasamy2013two,
hadizadeh2014three,
hadizadeh2014relativistic} 
to Tehran 
\cite{hadizadeh2007four,
bayegan2008three,
bayegan2008realistic,
bayegan2008low,
bayegan2009three,
hadizadeh2011solutions,
radin2017four} 
and Krak\'ow 
\cite{golak2010two,
glockle2010exact,
glockle20103n,
skibinski2010numerical,
skibinski2011recent,
golak2013three,
topolnicki2015first,
topolnicki2017operator,
topolnicki2017three} groups.

In Sec. \ref{Relativistic_interactions_formalism}, we present the 3D formalism for the relationship between relativistic and nonrelativistic $NN$ potentials. By projecting the quadratic relation between nonrelativistic and relativistic $NN$ potentials in momentum helicity basis states, we obtain the matrix elements of relativistic $NN$ potentials as a function of the magnitude of $2N$ relative momenta, the angle between them, and the helicity eigenvalues. We derive a single integral equation for $NN$ total spin state $s=0$ and four coupled integral equations for $s=1$.
In Sec. \ref{Relativistic_interactions_calculations}, we present our numerical results for the matrix elements of relativistic $NN$ potential obtained from CD-Bonn potential in different spin and isospin channels.
In Sec. \ref{numerical_tests}, we test the obtained relativistic potential by calculating and comparing deuteron binding energy and differential and total cross-sections of $np$ elastic scattering with corresponding nonrelativistic results.
Finally, a conclusion and outlook are provided in Sec. \ref{Conclusion}.

\section{Relativistic $NN$ potentials in a momentum helicity representation}
\label{Relativistic_interactions_formalism}

In this section, we show how to obtain the matrix elements of relativistic $NN$ interactions in a 3D scheme from the nonrelativistic interactions by solving a nonlinear equation derived by Kamada and Gl\"ockle. The relativistic interactions are designed to accurately reproduce the $NN$ bound and scattering observables.
To this aim and to check the accuracy of obtained relativistic interactions, as we show in Sec. \ref{numerical_tests}, one needs to solve the homogeneous and inhomogeneous LS integral equations \eqref{LS_deuteron}, \eqref{T_s0}, and \eqref{T_s1} in a momentum helicity representation to calculate relativistic deuteron binding energy and the scattering amplitudes to obtain the differential and total cross-sections. 

The relativistic and nonrelativistic $NN$ potentials, {\it i.e.}, $\Vr_r$ and $V_{nr}$, are related together by a quadratic operator equation as \cite{kamada2007realistic}
\begin{equation} \label{eq.V-v}
V_{nr}=\frac{1}{4m} \biggl(\omega(\hat{p})\Vr_r+ \Vr_r \omega(\hat{p})+\Vr_r^2 \biggr),
\end{equation}
where $m$ is the mass of nucleons, $p$ is the relative momentum of two nucleons ($\hat{p}$ is the operator), and $\omega(p)=2E(p)=2\sqrt{m^2+p^2}$.
To calculate the relativistic $NN$ potential $\Vr_r$ from a nonrelativistic potential $V_{nr}$ in a 3D representation, we present Eq. (\ref{eq.V-v}) in momentum helicity basis states.
The antisymmetrized momentum helicity basis states for a $2N$ system with total spin and isospin
$s$ and $t$, and the relative momentum $\bp$ are introduced as  \cite{fachruddin2000nucleon}
\begin{equation} \label{eq.basis_helicity_a}
\ketp  \equiv   \frac {1}{\sqrt{2}}
   \biggl (1-\eta_{\pi}(-)^{s+t} \biggr) \  |{\bf p};\bphat  \  s\lam \rangle_{\pi}\  |t\rangle ,
\end{equation}
where $\bphat$ is the unit momentum operator, $\lam$ is the eigenvalue of the helicity operator ${\bf
s\cdot\bphat } $, with the parity eigenvalues $\eta_{\pi}=\pm1$ and eigenstates
$|\bp;\bphat  \  s\lam \rangle_{\pi}=\frac{1}{\sqrt{2}} (1+\eta_{\pi} P_{\pi}) |\bp;\bphat  \  s\lam \rangle$.
The $2N$ helicity basis states are orthogonal and normalized as
\begin{eqnarray}
\label{eq.completeness}
&& \brappleftprime  \ketp  = \biggl (1-\eta_{\pi}(-)^{s+t}\biggr)\ \delta_{\eta_{\pi'} \eta_{\pi}}  \delta_{s's} \delta_{t't} \  
 \biggl \{\delta(\bp'-\bp)\delta_{\lam\lam'}+\eta_{\pi}(-)^{s}
\delta(\bp'+\bp)\delta_{\lam'-\lam} \biggr \}, \cr
&& \sum_{s\lam\pi t}\int d\bp\  \ketp  \ \frac{1}{4}\ \brap   =1.
\end{eqnarray}
The matrix elements of $NN$ nonrelativistic and relativistic potentials in $2N$ helicity basis states, introduced in Eq. (\ref{eq.basis_helicity_a}), are given as
\begin{eqnarray}
V^\pist_{nr,\lam\lam'}(\bp,\bp') &\equiv&
\brap  V_{nr} \ketpp, \\
\Vr^\pist_{r,\lam\lam'}(\bp,\bp') &\equiv&
\brap  \Vr_r\ketpp.
\label{eq.Vr_in_helicity}
\end{eqnarray}
Representation of the quadratic relation between nonrelativistic and relativistic $NN$ potentials, given in Eq. (\ref{eq.V-v}), in $2N$ helicity basis states is as
\begin{eqnarray}
\label{eq.4mV_in_helicity}
\brap 4m V_{nr} \ketpp &=&
\brap  \omega(\hat{p})\Vr_r\ketpp
\nonumber \\  &+&
\brap \Vr_r\omega(\hat{p})\ketpp
\nonumber \\  &+&
\brap \Vr_r \cdot \Vr_r\ketpp .
\end{eqnarray}
The first and the second terms on the right-hand side of Eq. (\ref{eq.4mV_in_helicity}) can be evaluated straightforwardly. For evaluation of the third term, the completeness relation of Eq. (\ref{eq.completeness}) should be inserted. By these considerations, Eq. (\ref{eq.4mV_in_helicity}) reads as
\begin{eqnarray}
\label{eq.4mV_in_helicity-revised2}
 \frac{4m V^\pist_{nr, \lam\lam'}(\bp,\bp')}{\omdp}
  &=&
\Vr^\pist_{r,\lam\lam'}(\bp,\bp')
+
 \frac{1}{ \omdp }  \frac{1}{4} \sum_{\lam''=-1}^{+1}\int d\bp'' \
\Vr^\pist_{r,\lam\lam''}(\bp,\bp'')
\Vr^\pist_{r,\lam''\lam'}(\bp'',\bp').
\end{eqnarray}
By considering the following properties of $NN$ potentials, one can obtain the negative helicity eigenvalue components of the potential from the positive ones as
\begin{eqnarray}
\label{eq.helicity_symmetry}
 V^\pist_{nr, -\lam\lam'}(\bp,\bp')  &=& \eta_\pi(-)^s \  V^\pist_{nr, \lam\lam'}(-\bp,\bp'),  \cr
 V^\pist_{nr, \lam-\lam'}(\bp,\bp')  &=& \eta_\pi(-)^s \  V^\pist_{nr, \lam\lam'}(\bp,-\bp').
\end{eqnarray}
The above relations are also valid for the relativistic potential $\Vr^\pist_{r,\lam\lam''}(\bp,\bp'')$.
By using the properties of Eq. (\ref{eq.helicity_symmetry}), one can show that the second term of Eq. (\ref{eq.4mV_in_helicity-revised2}), for $\lam''=-1$ and $\lam''=+1$ are equal together
\begin{eqnarray}
\label{eq.symmetry-Lambda}
\int d\bp'' \
\Vr^\pist_{r,\lam -1}(\bp,\bp'')
\Vr^\pist_{r,-1 \lam'}(\bp'',\bp')
 &=&
 \int d\bp'' \
\eta_\pi(-)^s \  \Vr^\pist_{r,\lam 1}(\bp,-\bp'')
\eta_\pi(-)^s \  \Vr^\pist_{r,1 \lam'}(-\bp'',\bp')
\nonumber \\
 &=&
\underbrace{\eta_\pi^2 (-)^{2s}}_1  \int d\bp'' \
  \Vr^\pist_{r,\lam 1}(\bp,-\bp'')
\Vr^\pist_{r,1 \lam'}(-\bp'',\bp')
\nonumber \\
 &=&
 \int d\bp'' \
  \Vr^\pist_{r,\lam 1}(\bp,\bp'')
\Vr^\pist_{r,1 \lam'}(\bp'',\bp').
\end{eqnarray}
For $2N$ singlet spin state, $s=0$, Eq. (\ref{eq.4mV_in_helicity-revised2}) leads to a single integral equation to obtain the matrix elements of relativistic potential $\Vr^\pizt_{r,00}(\bp,\bp')$
\begin{eqnarray}
\label{eq.4mV_in_helicity-s=0}
\frac{4m V^\pizt_{nr, 00}(\bp,\bp')}{\omdp} &=&
  \Vr^\pizt_{r,00}(\bp,\bp')
+
 \frac{1}{ \omdp }  \frac{1}{4} \int d\bp'' \
\Vr^\pizt_{r,00}(\bp,\bp'')
\Vr^\pizt_{r,00}(\bp'',\bp').
\end{eqnarray}
For triplet spin states, $s=1$, Eq. (\ref{eq.4mV_in_helicity-revised2}) leads to four coupled integral equations, corresponding to helicity eigenvalues $\lam,\lam'=0,+1$, to calculate the matrix elements of relativistic potentials $\Vr^\piot_{r,00}(\bp,\bp')$, $\Vr^\piot_{r,01}(\bp,\bp')$, $\Vr^\piot_{r,10}(\bp,\bp')$, and $\Vr^\piot_{r,11}(\bp,\bp')$
\begin{eqnarray}
  \label{eq.4mV_in_helicity-s=1}
   \frac{4m V^\piot_{nr, \lam\lam'}(\bp,\bp')}{\omdp} &=&
 \Vr^\piot_{r,\lam\lam'}(\bp,\bp')
 \cr &+&
 \frac{1}{ \omdp }  \biggl ( \frac{1}{2} \int d\bp'' \
\Vr^\piot_{r,\lam1}(\bp,\bp'')
\Vr^\piot_{r,1\lam'}(\bp'',\bp')
+\frac{1}{4} \int d\bp'' \
\Vr^\piot_{r,\lam0}(\bp,\bp'')
\Vr^\piot_{r,0\lam'}(\bp'',\bp') \biggr) .
\end{eqnarray}
Eqs. (\ref{eq.4mV_in_helicity-s=0}) and (\ref{eq.4mV_in_helicity-s=1}) should be solved for each value of $2N$ total isospin $t=0, 1$.
For the numerical solution of single and coupled integral equations, {\it i.e.}, Eqs. (\ref{eq.4mV_in_helicity-s=0}) and (\ref{eq.4mV_in_helicity-s=1}), by choosing momentum vector $\bpp$ parallel to the $z-$axis, the azimuthal angular dependence of the matrix elements of nonrelativistic and relativistic potentials can be factored out as an exponential phase as
\begin{eqnarray}
V^\pist_{nr, \lam\lam'}(\bp,p'\hat{z}) &=&  e^{{i\lam' \phi} } V^\pist_{nr, \lam\lam'}(p,p',x) ,  \cr
\Vr^\pist_{r,\lam\lam'}(\bp,p'\hat{z}) &=&  e^{{i\lam' \phi} } \Vr^\pist_{r,\lam\lam'}(p,p',x) ,  \cr
\Vr^\pist_{r,\lam'' \lam'}(\bpz,p'\hat{z}) &=&  e^{{i\lam' \phi''} }   \Vr^\pist_{r,\lam'' \lam'}(p'',p',x''),
\end{eqnarray}
where $x \equiv \bphat \cdot \bphat'$,  $x'' \equiv \bphat'' \cdot \bphat'$.
One can show that the matrix elements of $\Vr^\pist_{r,\lam \lam''} (\bp,\bp'')$ can be obtained as
\begin{equation}
\label{V_p_pz_vector}
\Vr^\pist_{r,\lam \lam''}  (\bp,\bp'' )
=
\frac{\Vr^\pist_{r,\lam\lam''}(p,p'',\gamma)}{d_{\lam'' \lam}^s (\gamma)} \sum_{\lam^*=-s}^{+s} e^{i\lam^*(\phi-\phi'')}  d_{\lam^* \lam} ^s (x) \   d_{\lam^* \lam''} ^s (x'')  ,
\end{equation}
where $\gamma = x x'' + \sqrt{1-x^2} \sqrt{1-x''^2} \cos{(\phi-\phi'')}$.
By these considerations, Eqs. (\ref{eq.4mV_in_helicity-s=0}) and (\ref{eq.4mV_in_helicity-s=1}) can be written as
\begin{eqnarray}
  \label{eq.4mV_in_helicity-s=0_revised2}
   \frac{4m V^\pizt_{nr, 00}(p,p',x)}{\omdp} &=&
 \Vr^\pizt_{r,00}(p,p',x)
+
 \frac{1}{ \omdp }   \frac{1}{4}  \int _{0}^{\infty} dp'' p''^2 \int _{-1}^{+1} dx'' \
\Vr^{\pizt, 0}_{r,0 0}  (p,p'',x,x'')
 \      \Vr^\pizt_{r, 0 0}(p'',p',x'') ,  \quad\quad \\ \cr
   \label{eq.4mV_in_helicity-s=1_revised2}
   \frac{4m V^\piot_{nr, \lam\lam'}(p,p',x)}{\omdp} &=&
 \Vr^\piot_{r,\lam\lam'}(p,p',x)
 \cr &+&
 \frac{1}{ \omdp }    \int _{0}^{\infty} dp'' p''^2 \int _{-1}^{+1} dx'' \cr
 &\times&   \biggl \{
 \frac{1}{4} \Vr^{\piotlp}_{r,\lam 0}  (p,p'',x,x'')
 \      \Vr^\piot_{r, 0 \lam'}(p'',p',x'') 
 + 
 \frac{1}{2} \Vr^{\piotlp}_{r,\lam 1}  (p,p'',x,x'')
 \     \Vr^\piot_{r, 1 \lam'}(p'',p',x'') \biggr \} , 
\end{eqnarray}
where
\begin{eqnarray}
 \Vr^{\pistlp}_{r,\lam \lam''}  (p,p'',x,x'') &=&
\sum_{\lam^*=-s}^{+s}   d_{\lam^* \lam} ^s (x) \   d_{\lam^* \lam''} ^s (x'')
 \int_0^{2\pi} d\phi'' \  e^{i (\lam^*- \lam' ) (\phi-\phi'')}
\frac{ \Vr^\pist_{r,\lam\lam''}(p,p'',\gamma) }{d_{\lam'' \lam}^s (\gamma)}.
\label{phi-integration}
\end{eqnarray}
The $\phi''$ integration in the interval $[0,2\pi]$ can be simplified to $[0,\pi/2]$ interval by
\begin{eqnarray}
 \int_0^{2\pi} d\phi'' \  e^{im (\phi-\phi'')} {\cal F}\bigl(\cos(\phi-\phi'')\bigr)&=&\int_0^{2\pi} d\phi'' \  e^{im \phi''} {\cal F}\bigl(\cos \phi''\bigr)\cr &=& 2
   \int_0^{\pi/2} d\phi''   \cos m\phi'' \biggl [ {\cal F}\bigl(\cos \phi'' \bigr)  + (-)^m {\cal F}\bigl(-\cos \phi'' \bigr)   \biggr] .
\label{phi-integration_simplification}
\end{eqnarray}
%

\section{Calculation of Relativistic $NN$ Interactions}
\label{Relativistic_interactions_calculations}

\begin{figure}[H]
\centering
  \includegraphics[width=0.95\columnwidth]{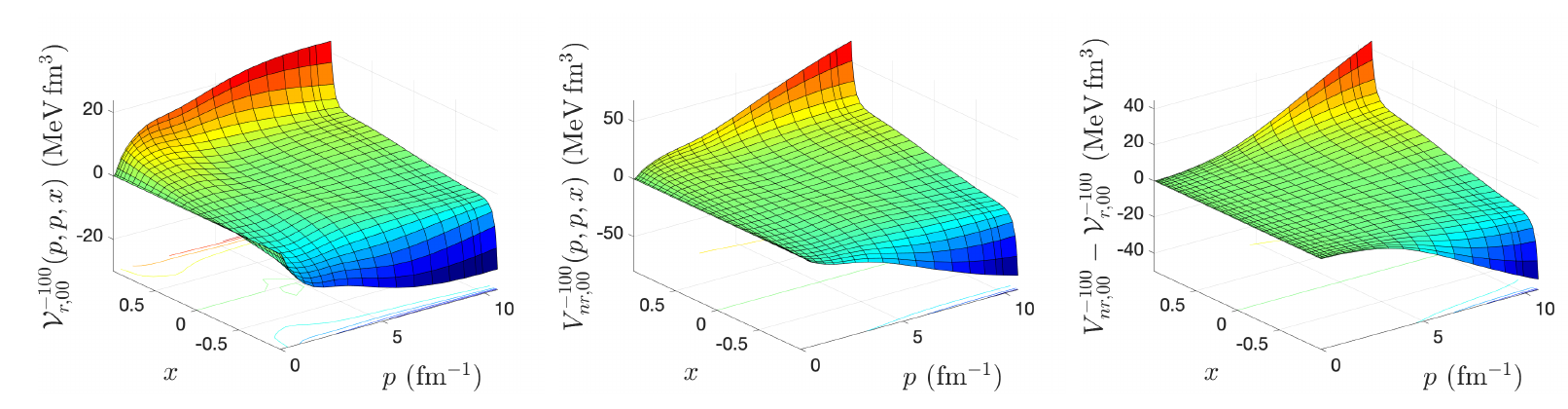}
\caption{The matrix elements of the relativistic $NN$ potential (left panel), the nonrelativistic $NN$ potential (middle panel), and their differences (right panel), calculated for CD-Bonn potential in $(s=0, \ t=0)$ channel, as a function of $2N$ relative momenta $p=p'$ and the cosine of the angle between them $x$.}
\label{fig_px-s0t0}
\end{figure}

\begin{figure}[H]
\centering
  \includegraphics[width=0.95\columnwidth]{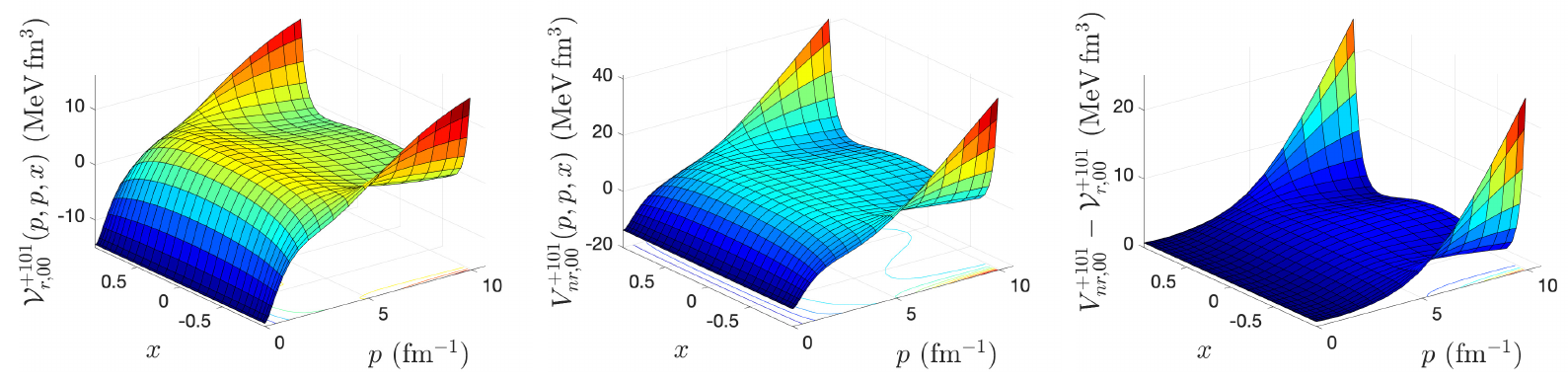}
\caption{The same as Fig. \ref{fig_px-s0t0}, but for $(s=0, \ t=1)$ channel.}
\label{fig_px-s0t1}
\end{figure}

\begin{figure}[H]
\centering
  \includegraphics[width=0.95\columnwidth]{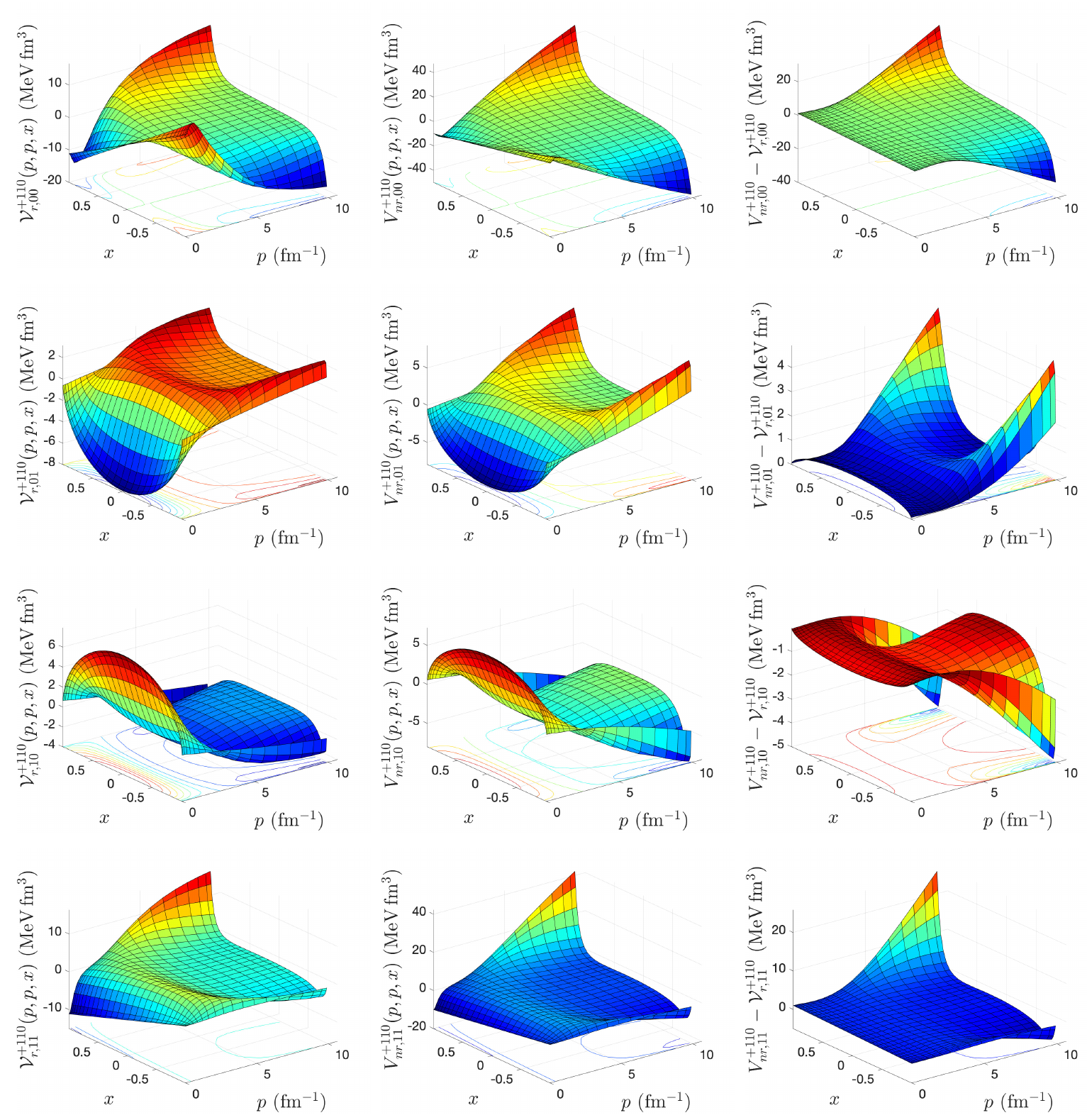}
\caption{The same as Fig. \ref{fig_px-s0t0}, but for $(s=1, \ t=0)$ channel.}
\label{fig_px-s1t0}
\end{figure}

\begin{figure}[H]
\centering
  \includegraphics[width=0.95\columnwidth]{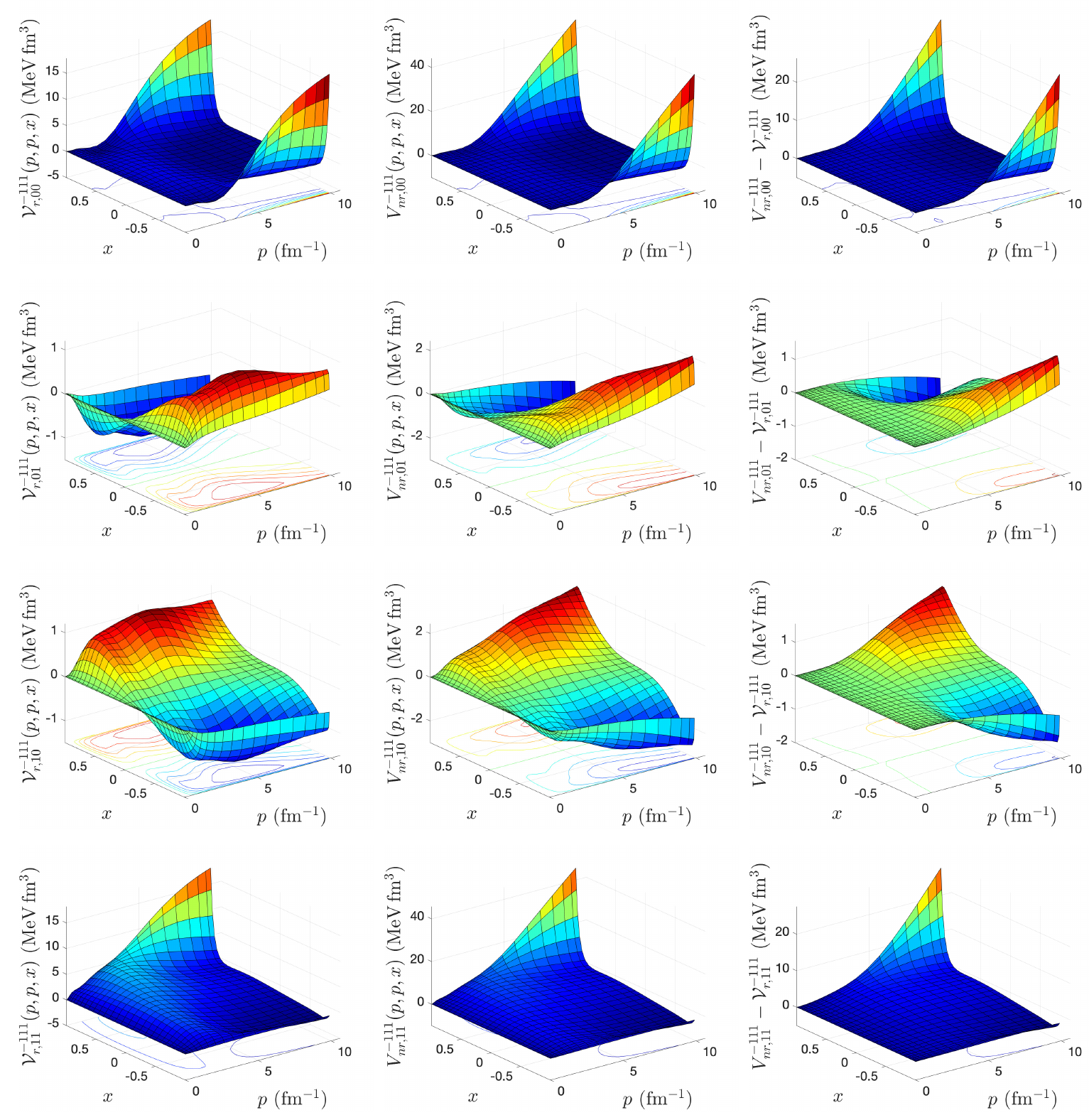}
\caption{The same as Fig. \ref{fig_px-s0t0}, but for $(s=1, \ t=1)$ channel.}
\label{fig_px-s1t1}
\end{figure}

To calculate the matrix elements of relativistic potentials for different spin-isospin $(s,t)$ channels, we solve the integral equations (\ref{eq.4mV_in_helicity-s=0_revised2}) and (\ref{eq.4mV_in_helicity-s=1_revised2}) using an iterative method proposed by Kamada and Gl\"ockle \cite{kamada2007realistic}. The iteration starts with $\Vr^{(0)\pist}_{r,\lam\lam'}(p,p',x) = \dfrac{4m V^\pist_{nr, \lam \lam'}(p,p',x)}{\omdp}$ and stops when the maximal difference between the matrix elements of the relativistic potential $\Vr^{\pist}_{r,\lam \lam'}(p,p',x)$ obtained from two successive iterations drops below $10^{-6}$ MeV\ fm$^3$.
After each iteration, to obtain the matrix elements $\Vr^{\pistlp}_{r,\lam \lam''} (p,p'',x,x'')$ that appears in the kernel of Eqs. (\ref{eq.4mV_in_helicity-s=0_revised2}) and (\ref{eq.4mV_in_helicity-s=1_revised2}), we need to perform the azimuthal angle integration of Eq. (\ref{phi-integration}).
To speed up the convergence of the iteration in solving Eqs. (\ref{eq.4mV_in_helicity-s=0_revised2}) and (\ref{eq.4mV_in_helicity-s=1_revised2}), in some $(s,t)$ channels even to be able to reach the convergence, we use the weighted average of relativistic potential obtained from two successive iterations as
\begin{eqnarray} \label{eq.convergence-factor}
 \Vr_r^{(n)}  \longrightarrow \frac{\alpha
\Vr_r^{(n)} + \beta \Vr_r^{(n-1)} }{\alpha+\beta}; \ n=1,2,...
\end{eqnarray}
The input to our calculations is a 3D form of CD-Bonn potential in momentum helicity representation obtained from the summation of partial wave matrix elements of the potential up to total angular momentum $j_{max} = 20$. For the discretization of the continuous momentum and angle variables, we use the Gauss-Legendre quadrature. A combination of hyperbolic and linear mapping with 120 mesh points is used for the momentum variables, and for azimuthal and polar angle variables, a linear mapping with 40 mesh points is used. 
In our calculations, we use the nucleon mass $m = \frac{2 m_p m_n}{ m_p + m_n} = 938.91852$ MeV and the conversion factor~$ \hslash c = 197.327053 \ \text{MeV fm} = 1$, where proton and neutron masses are $m_p = 938.27231$ MeV and $m_n = 939.56563$ MeV \cite{machleidt2001high}.
The number of iterations needed to reach convergence in the matrix elements of relativistic potential calculations in different spin-isospin channels, for different values of the weight averaging parameters $\alpha$ and $\beta$, is shown in Table \ref{table_convergence-alpha-beta}.
Our numerical analysis indicates that for $(s=0, \ t=0)$ channel, the convergence can be reached only for $\alpha=\beta=1$, whereas for $(s=0, \ t=1)$ and $(s=1, \ t=0)$ the fastest convergence can be reached by $\alpha=2, \ \beta=1$. For $(s=1, \ t=1)$ the fastest convergence can be reached by $\alpha=3, \ \beta=1$.
It should be mentioned that Kamada and Gl\"ockle have used $\alpha = \beta = 1$ in their calculations to obtain relativistic potentials from AV18, CD-Bonn, and Nijm I, II potentials \cite{kamada2007realistic}.
%
\begin{table}[htp!]
\caption{The number of iterations $N_{iter}$ needed to reach convergence in Eqs. (\ref{eq.4mV_in_helicity-s=0_revised2}) and (\ref{eq.4mV_in_helicity-s=1_revised2}) for the calculation of relativistic potential, obtained from CD-Bonn potential, as a function of the weight averaging parameters $\alpha$ and $\beta$, which are defined in Eq. (\ref{eq.convergence-factor}).}
\centering
\begin{tabular}{ccccccccccccc}
\hline
 $\alpha$ &&  $\beta$ && \multicolumn{7}{c}{$N_{iter}$}
   \\  \cline{5-11}
   && && $(s=0, \ t=0)$ && $(s=0, \ t=1)$ && $(s=1, \ t=0)$ && $(s=1,\ t=1)$  \\
\hline
1 &&   0  &&  -   &&  -    &&   -  &&  27 \\
1 &&   1  &&  76   &&  18    && 21    &&  19 \\
2 &&   1  &&  -   &&  13    && 21    &&  12  \\
3 &&   1  &&  -   &&   19   &&  34   && 10  \\
4 &&   1  &&   -  &&  25    &&  50   && 12  \\
5 &&   1  &&   -  &&   30   &&  69   &&  14 \\
\hline
\end{tabular}
\label{table_convergence-alpha-beta}
\end{table}
%
%
In Figs. \ref{fig_px-s0t0}-\ref{fig_px-s1t1}, the matrix elements of relativistic potential obtained from CD-Bonn potential in different spin-isospin channels are compared with corresponding nonrelativistic potentials. The differences between nonrelativistic and relativistic potentials are also shown.  
As we can see, while the nonrelativistic and relativistic potentials show similar structures, the difference between them is significant and is in the same order of magnitude as the potentials.
%

\section{Numerical tests for the relativistic $NN$ potentials }
\label{numerical_tests}

In this section, we present two numerical tests for NN bound and scattering states, which show the validity of our formalism and the accuracy of calculated relativistic potentials in the 3D scheme.  
\subsection{Deuteron binding energy and wave function}
To test the accuracy of the calculated relativistic $NN$ potential in $(s=1,\ t=0)$ channel, we calculate the deuteron binding energy and wave function for both nonrelativistic CD-Bonn and the obtained relativistic potentials.
The relativistic form of the homogenous LS equation for describing deuteron binding energy $E_d = m_d - 2m$ and wave function $\psi_{M_d}(p)$ is given by the following coupled integral equations for wave function components $\psi_0$ and $\psi_1$
\begin{equation}
 \psi_{M_{d}}(p) =  \frac{2\pi}{ m_d - \omega(p) }  \int _{0}^{\infty}dp'\  p'^{2}
 \left\{ \frac{1}{2}
 \Vr_{r,M_{d} 1 }^{+110}(p,p')
 \psi_{1}(p') +
 \frac{1}{4}
 \Vr_{r,M_{d}0}^{+110}(p,p')
 \psi_{0}(p')\right\}  , \label{LS_deuteron}
\end{equation}
where
\begin{equation}
\Vr_{r,M_{d}\Lambda'}^{+110}(p,p')  = \int _{-1}^{1}dx\  \Vr_{r,M_{d}\Lambda'}^{+110}(p,p',x) \  d^{1}_{M_{d}\Lambda'}(x) . \label{label28}
\end{equation}

\begin{figure}[hbt!]
\centering
  \includegraphics[width=0.6\columnwidth]{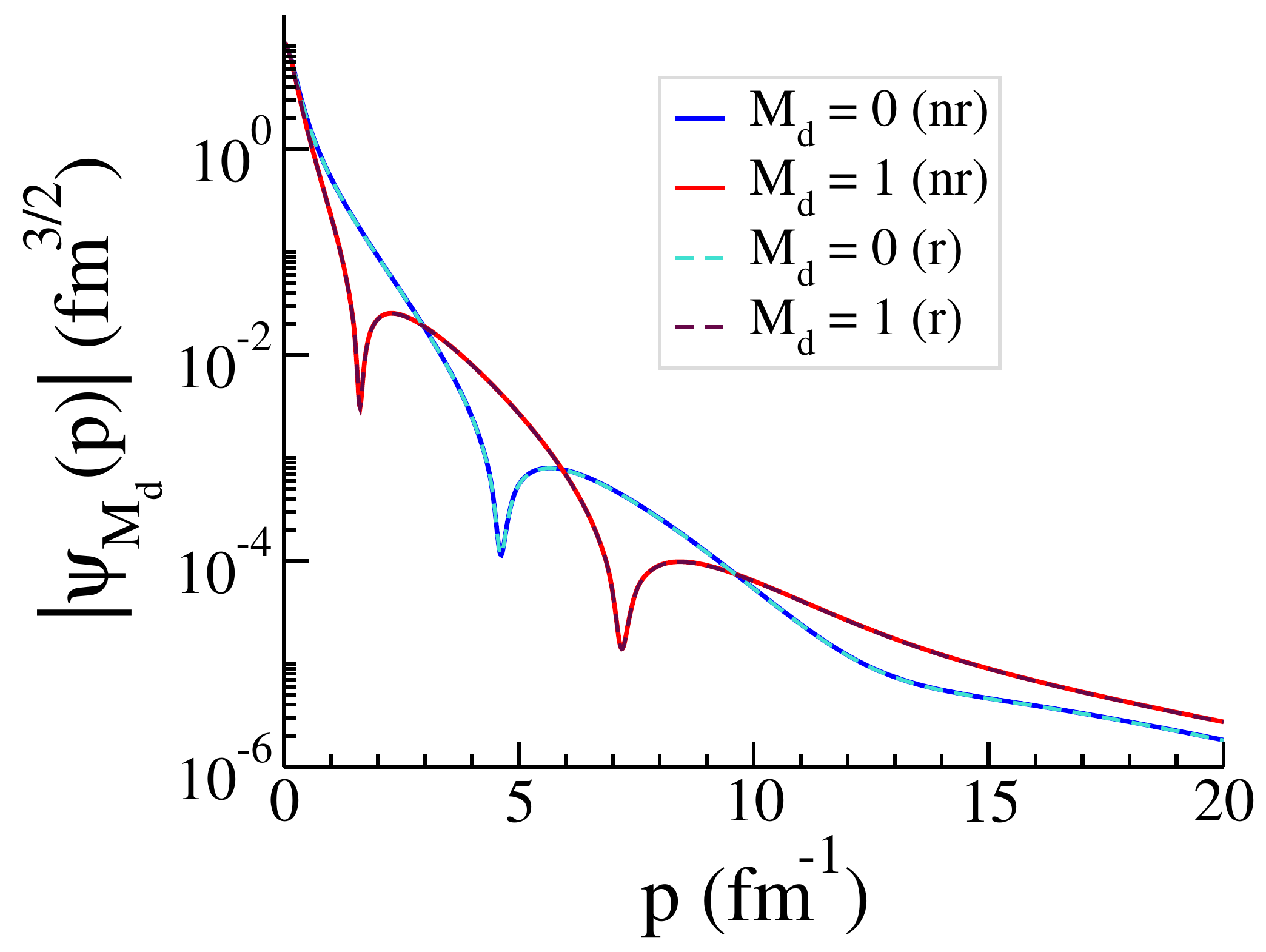}
\caption{The absolute value of deuteron wave function components $\psi_0$ and $\psi_1$, calculated for relativistic (r) and nonrelativistic (nr) CD-Bonn potentials, as a function of the relative momentum $p$.}
\label{fig_deuteron}
\end{figure}

In the nonrelativistic form, the free propagator is replaced by $(E_d - \frac{p^{2}}{m})^{-1}$ and the relativistic $NN$ potential is replaced by the nonrelativistic potential $V_{nr,M_{d}\Lambda'}^{+110}(p,p')$ \cite{fachruddin2001new}.
In Table \ref{table_deuteron}, we present our numerical results for deuteron binding energies obtained from both CD-Bonn and relativistic potentials. The relative percentage difference of $0.06 \ \%$ indicates an excellent agreement between relativistic and nonrelativistic deuteron binding energies. 
In Fig. \ref{fig_deuteron}, we show the deuteron wave function components calculated for both relativistic and nonrelativistic CD-Bonn potentials. 
As we can see, the constructed relativistic potential reproduces the deuteron binding energy and wave function obtained by CD-Bonn potential with high accuracy.

\begin{table}[hbt!]
\caption{Calculated deuteron binding energy for relativistic and nonrelativistic CD-Bonn potentials and their relative percentage difference.}
\centering
\begin{tabular}{ccccccc}
\hline
 $E_d^{r}$ (MeV) &&  $E_d^{nr}$ (MeV) &&  $ \left | \dfrac{(E_d^{r}-E_d^{nr})}{E_d^{nr}} \right  | \times 100  \%$
   \\  \hline
$-2.22463$ && $-2.22325$    &&   $0.06207$ \\
\hline
\end{tabular}
\label{table_deuteron}
\end{table}

\subsection{$np$ elastic scattering}

\begin{figure}[hbt!]
\centering
  \includegraphics[width=0.95\columnwidth]{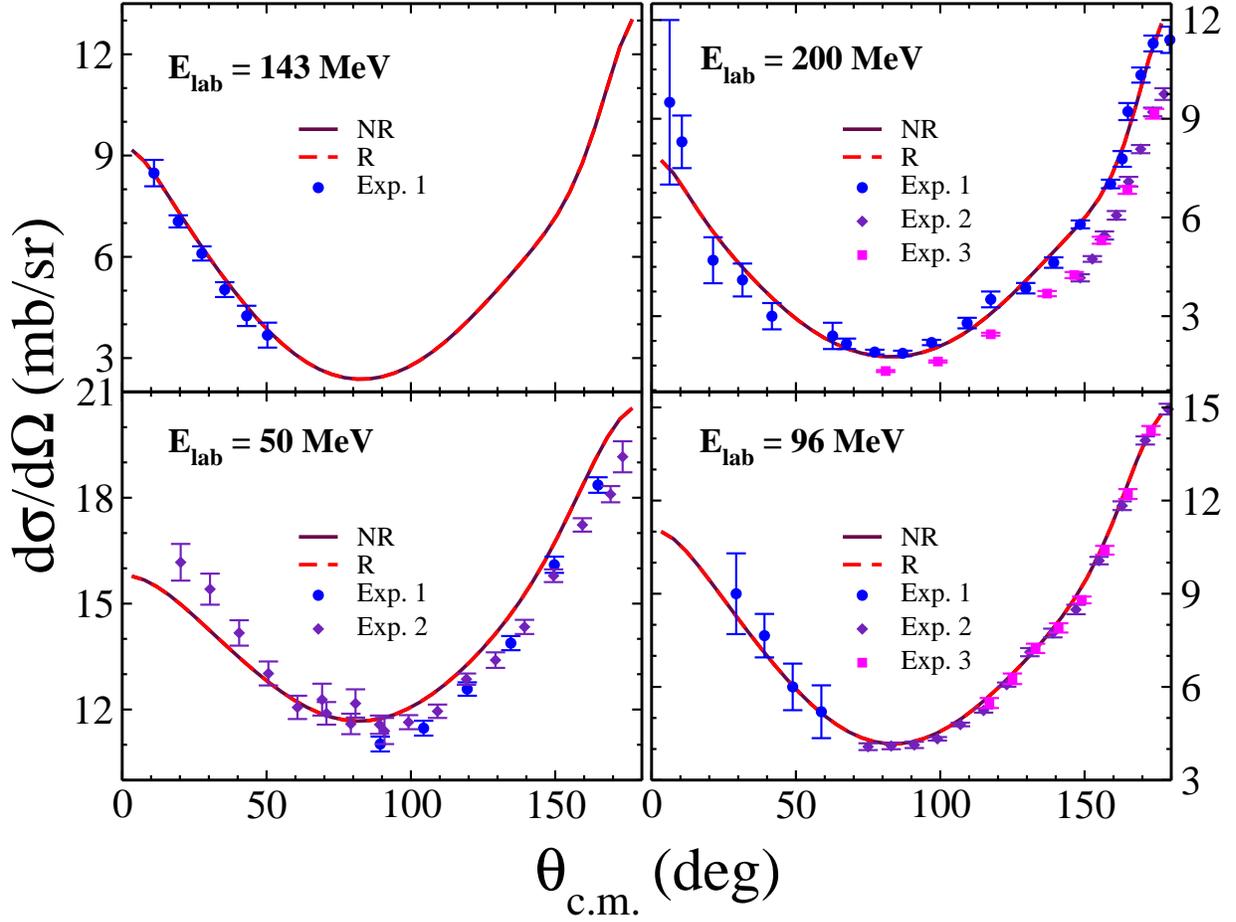}
\caption{The $np$ differential cross sections calculated for relativistic (solid magenta lines) and nonrelativistic CD-Bonn (dashed red lines) potentials as a function of scattering angle $\theta_{c.m.}$ in the CM frame, for the incident projectile energies $E_{lab}=50, \ 96, \ 143$, and $200$ MeV. The experimental data are from Refs. \cite{Fink_NPA518,Montgomery_PRC16} for $E_{lab} = 50$ MeV (Exp.~1 and Exp.~2, respectively), form Refs. \cite{Griffith_PPS71,Rahm_PRC63,Ronnqvist_PRC45} for $E_{lab} = 96$ MeV (Exp.~1, Exp.~2 and Exp.~3, respectively), from Ref. \cite{Bersbach_PRD13} for $E_{lab} = 142.8$ MeV (Exp. 1) and from Refs. \cite{Hurster_PLB90, Kazarinov_SPJ16, Franz_PST87} for $E_{lab} = 200$ MeV (Exp.~1, Exp.~2 and Exp.~3, respectively).}
\label{fig_dsigma}
\end{figure}

For the second numerical test, we calculate the differential and total cross-section of $np$ elastic scattering for the relativistic potential constructed from CD-Bonn potential. To describe the relativistic $np$ elastic scattering in momentum helicity space, the relativistic form of inhomogeneous LS equations for $2N$ $t-$matrices in singlet and triplet spin states can be obtained as
\begin{eqnarray}
T^\pizt_{00}(p,p',x) &=& \Vr^\pizt_{r, 00}(p,p',x) \cr
&+& \frac{1}{4}\int_{0}^{\infty}dp''p''^2\int_{-1}^{+1}dx''\ \frac{\Vr^\piztz_{r,0 0}  (p,p'',x,x'')\ T^\pizt_{00}(p'',p',x'')}{\omega(p_{0})-\omega(p'')+i\epsilon}, \label{T_s0}
\\ \cr 
T^\piot_{\lam\lam'}(p,p',x) &=& \Vr^\piot_{r, \lam\lam'}(p,p',x) \cr
&+& \frac{1}{2} \int_{0}^{\infty}dp''p''^2 \int_{-1}^{+1} dx''\ \frac{\Vr^\piotlp_{r,\lam1}  (p,p'',x,x'')\ T^\piot_{1\lam'}(p'',p',x'')}{\omega(p_{0})-\omega(p'')+i\epsilon} \cr
&+& \frac{1}{4}\int_{0}^{\infty} dp'' p''^2 \int_{-1}^{+1} dx'' \ \frac{\Vr^\piotlp_{r,\lam0}  (p,p'',x,x'')\ T^\piot_{0\lam'}(p'',p',x'')}{\omega(p_{0})-\omega(p'')+i\epsilon}.
 \label{T_s1}
\end{eqnarray}
The on-shell momentum $p_0$ for an incident projectile energy $E_{lab}$ is defined by $p_0 = \sqrt{ \frac{ m_p^2 E_{lab} (E_{lab} + 2m_n ) }{ (m_p + m_n)^2 + 2 m_p E_{lab} } }$ \cite{machleidt2001high}.
By replacing the relativistic free propagator $(\omega(p_{0})-\omega(p'')+i\epsilon)^{-1}$ with
$(\frac{p_{0}^{2}}{m} - \frac{p''^{2}}{m} +i\epsilon)^{-1}$ and relativistic potential $\Vr_{r}$ with the nonrelativistic potential $V_{nr}$, Eqs. (\ref{T_s0}) and (\ref{T_s1}) yield the nonrelativistic form of LS equations of Ref. \cite{fachruddin2000nucleon}.
The relativistic differential cross section of $np$ elastic scattering as a function of the incident projectile energy is given by
\begin{eqnarray}
\frac{d\sigma}{d\Omega} = (2\pi)^{4} \left (\frac{\omega(p_{0})}{4} \right)^{2}\frac{1}{4} \sum_{m'_{s_{1}}m'_{s_{2}}m_{s_{1}}m_{s_{2}}} \left |T^{phys}_{m'_{s_{1}}m'_{s_{2}}m_{s_{1}}m_{s_{2}}}(p_{0},p_{0},x') \right|^{2} ,
\end{eqnarray}
where the matrix elements of the on-shell physical $t-$matrices can be obtained from the on-shell relativistic $NN$ $t-$matrices of Eqs. (\ref{T_s0}) and (\ref{T_s1}) as
\begin{eqnarray}
T^{phys}_{m'_{s_{1}}m'_{s_{2}}m_{s_{1}}m_{s_{2}}}(p_{0},p_{0},x) &=& 
\frac{1}{4}\sum_{\pist}(1-\eta_{\pi}(-)^{s+t})\ 
C^2 \left (\frac{1}{2}\frac{1}{2}t,m_{t_{1}}m_{t_{2}} \right )\cr 
&\times&\  \ C \left (\frac{1}{2}\frac{1}{2}s,m'_{s_{1}}m'_{s_{2}}\lambda'_{0} \right) \ 
C \left (\frac{1}{2}\frac{1}{2}s,m_{s_{1}}m_{s_{2}}\lambda_{0} \right) 
 \  \sum_{\lambda'}d^{s}_{\lambda'_{0}\lambda'}(x)\ T^{\pist}_{\lambda'\lambda_{0}}(p_{0},p_{0},x),
\end{eqnarray}
where $m_{s_{i}}$ and $m_{t_{i}}$ are the spin and isospin projection of single nucleons along the quantization $z-$axis, and the coefficients $C$ are the Clebsch-Gordan coefficients.
In Fig. \ref{fig_dsigma}, we show the differential cross sections of $np$ elastic scattering, calculated for relativistic and nonrelativistic CD-Bonn potentials, for the projectile energies $E_{lab}=50, 96, 143$ and $200$ MeV. 
Total cross-sections can provide a more detailed comparison between relativistic and nonrelativistic potentials. In Table \ref{table_sigma}, we present our numerical results for total cross-sections of $np$ elastic scattering, obtained by relativistic and nonrelativistic CD-Bonn potentials, as a function of the incident projectile energy $E_{lab}$. As we can see, the maximum relative percentage difference is less than $0.01 \ \%$, indicating that the relativistic total cross-sections are in excellent agreement with the corresponding nonrelativistic cross-sections.

\begin{table}[hbt!]
\caption{The total cross section of $np$ elastic scattering as a function of the incident projectile energy $E_{lab}$, calculated for relativistic ($\sigma_{r}$) and nonrelativistic ($\sigma_{nr}$) CD-Bonn potentials.} 
\centering
\begin{tabular}{ccccccccccccc}
\hline
 $E_{lab}$ (MeV)  &   $\sigma_{r}$ (mb) & $\sigma_{nr}$ (mb) & $\left | \frac{\sigma_{r} - \sigma_{nr}}{\sigma_{nr}} \right | \times 100 \%$ \\
\hline
0.001 &  20337.1 & 20336.6 &  0.00246    \\
0.01  &  19222.1 & 19221.7 &   0.00208  \\
0.1   &  12786.7 &  12786.6 &  0.00078  \\
1     &  4262.94 & 4262.95 & 0.00023  \\
3     &  2291.39 &  2291.40 & 0.00044  \\
5     &  1632.54 & 1632.55 &   0.00061 \\
10    &  943.252 & 943.253 &  0.00011  \\
20    &  483.491 & 483.491 &  0.00000   \\
50    &  167.926 & 167.925 &  0.00060    \\
100   &  75.0321 & 75.0359 & 0.00506    \\
250   &  37.6465 & 37.6485 &  0.00531   \\
500   &  28.2841 & 28.2869 & 0.00990     \\
750   &  24.4850 & 24.4872 &  0.00898   \\
\hline
\end{tabular}
\label{table_sigma}
\end{table}

Kamada and Gl\"ockle have shown in Ref. \cite{kamada2007realistic} that the obtained relativistic potential from AV18 potential, in a PW decomposition, reproduces the nonrelativistic phase shifts with five significant figures with projectile energy in the domain $(1-350)$ MeV. As one can see in Table \ref{table_sigma}, our nonrelativistic and relativistic cross-section results are also in perfect agreement with five significant figures with an incident projectile energy in the broader domain $(0.001-750)$ MeV. So we are convinced that the 3D formulation and calculations for relativistic potential provide the same accuracy as a PW calculation.
 
Moreover, in a prior study for calculation of relativistic potentials from spin-independent MT potential \cite{Hadizadeh2017}, which has no spin and isospin complexity of CD-Bonn, we obtained a relative percentage difference of $0.06\%$ between nonrelativistic and relativistic deuteron binding energies and a maximum relative percentage difference of $0.007\%$ in two-body total elastic scattering cross-sections, which can be compared with $0.06\%$ and $0.01\%$ relative percentage differences obtained in this study for CD-Bonn potential. This comparison indicates calculating relativistic $NN$ interactions from realistic interactions in a 3D scheme provides almost the same accuracy level as a spin-independent calculation.

\section{Conclusion and outlook}
\label{Conclusion}
In this paper, the quadratic equation, which connects the relativistic and nonrelativistic $NN$ potentials, is formulated in momentum helicity space as a single and four coupled three-dimensional integral equations for $2N$ singlet and triplet spin states. 
In our numerical calculations, we implement the CD-Bonn potential to obtain the matrix elements of the relativistic potential as a function of the magnitude of $2N$ relative momenta, the angle between them, and spin and isospin quantum numbers. 
The quadratic integral equations are solved using an iterative scheme. Our numerical results indicate that calculated relativistic $NN$ potential from the CD-Bonn potential reproduces $2N$ observables for deuteron binding energy and the differential and total cross sections of $np$ elastic scattering with high accuracy.
The implementation of relativistic $NN$ potentials in the relativistic description of triton binding energy and wave function is currently underway.

\bibliography{references}

\section*{Acknowledgements}
The work of M. R. Hadizadeh was supported by the National Science Foundation under grant No. NSF-PHY-2000029 with Central State University.

\section*{Author contributions statement}
M.R.H. designed and directed the project; M.R.H. and M.R. developed the theoretical formalism; F.N. calculated the matrix elements of CD-Bonn potential in a 3D scheme; M.R.H. and M.R. calculated the matrix elements of relativistic potentials; all authors discussed the results and contributed to the final manuscript.

\section*{Competing financial interests}
The authors declare no competing interests.

\section*{Data Availability}
The data that support the findings of this study are available from the corresponding author upon reasonable request.

\end{document}